# Direct metal nano-imprinting using embossed solid electrolyte stamp

A Kumar[1], K H Hsu[2], K E Jacobs[2], P M Ferreira[2] and N X Fang[2,3]

[1]Electrical and Computer Engineering, University of Illinois at Urbana-Champaign, 1406 W Green Street, Urbana, IL 61801 USA
[2]Mechanical Science and Engineering, University of Illinois at Urbana-Champaign, 1206 W Green Street, Urbana, IL 61801 USA
[3]Mechanical Engineering, Massachusetts Institute of Technology, 77 Massachusetts Avenue, Room 3-173, Cambridge, MA, USA 02139

Email: anilku2@illinois.edu

**Abstract.** In this paper, we report direct patterning of metal nanostructures using an embossed solid electrochemical stamp. Microforming of solid superionic stamps using Si templates—analogous to polymer patterning in nano-imprint lithography—is explored. Silver sulfide ($Ag_2S$)—a superionic conductor with excellent micro-forming properties—is investigated as a candidate material. Important parameters of the superionic stamp, including mechanical behavior, material flow during forming, and feature recovery after embossing are studied. Excellent feature transferability during embossing as well as etching is observed. To illustrate the capability of this approach silver nano-antennas with gaps <10 nm were successfully fabricated. The possibility for large area patterning with stamp diameters >6 mm is also demonstrated. Embossing based metal patterning allows fabrication beyond two-dimensional nanofabrication and several patterning schemes are reported.

## 1. Introduction

Recent growth in miniaturization at nanoscale has been fueled by impressive progress in the ability to fabricate ever-smaller features. This has boosted progress in various fields including nano-photonics, optoelectronics, biological and chemical sensing, energy harvesting, diffraction-free lithography, and micro-and-nano manufacturing systems [1]. Traditionally, the indirect approach of nanofabrication using polymer patterning, and subsequent metal evaporation and lift-off process is employed [2]. This methodology is inherently complex and involves several steps before final pattern is generated. It is also highly sensitive to ambient conditions and requires significant capital and energy investment. Therefore, direct metal patterning methods can be a boon for next generation of micro- and nano-scale manufacturing because they require fewer steps, are carried out at ambient conditions, and do not demand serious input in terms of time and infrastructure [3]. We have recently reported one such direct process [4-5] that can fabricate metallic patterns at ambient conditions using nominal external potential and applied pressure. This Solid-State Superionic Stamping (S4) process, with stamps patterned using Focused Ion Beam (FIB) milling, has been successfully demonstrated for Ag [4] and Cu [5], the two most important metals for plasmonics and microelectronics, respectively.

Because FIB milling is limited in its ability to fabricate sub-50 nm features while writing over large areas, the true capability of S4 process in terms of smallest feature size and largest area is constrained by ability to pattern the superionic stamp surface. Additionally, due to limited life span of the stamp surface, a new milling step is required after certain number of etches, which dramatically increases cost and time to carry out the process. A micro-forming based approach of stamp fabrication—similar to nano-imprint lithography—can overcome these limitations. However, the stamp material needs to be conducive to such an embossing procedure and various parameters related to mechanical deformation of the stamp need to be investigated. Here we explore this possibility and report how combination of the best merits of nano-imprint lithography and micro-forming incorporated into the S4 process allows development of a new cost and energy efficient process. We demonstrate that this approach can directly fabricate metal nanostructures that are comparable or better than electron beam lithography but carried out at ambient conditions and at fraction of cost of indirect patterning processes. Using silver sulfide ($Ag_2S$) as a stamp material, we characterize various parameters of the stamp and fabrication process to demonstrate how this capability is achieved.

## 2. Stamp characterization
*2.1 Nano-indentation*

Silver sulfide is a ductile superionic conductor [6] with relatively low yield strength of 80 MPa [7]. Therefore, it is amenable to low pressure microforming and provides better conformability during electrochemical etching. To quantify the local deformation characteristics nanoindentation of $Ag_2S$ stamp surface was carried out. First, an $Ag_2S$ crystal was grown as reported earlier [4]. One end of this crystal was trimmed using microtoming which resulted into a flat mesa with surface roughness < 3 nm. This surface was used for nanoindentation using a Berkovich tip, an indent of which is shown in figure 1 (inset). The measurements were carried out in continuous stiffness mode (CSM) [8] using an Agilent G200 nano-indenter to incorporate dynamic effects during embossing. Reduced elastic modulus of approximately 30 GPa and hardness of 0.6 GPa was observed as seen in figure 1. These parameters for some of the commonly used material in device processing, including several metals and polymers with comparable mechanical properties are listed in Table 1.

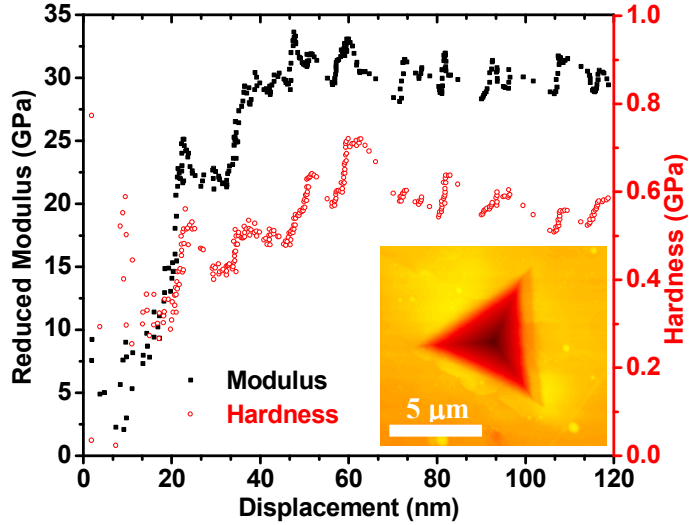

**Figure 1.** Nano-indentation of silver sulfide (Ag$_2$S) surface to investigate the elastic modulus and hardness of S4 stamps. The oscillations are attributed to dislocation-bursts to accommodate the stress [9, 10] which can be observed in the AFM image of the indent mark (inset).

A comparison in Table 1 shows Ag$_2$S has much lower hardness compared to Si and quartz which are the two most common material used for template design. Therefore, standard templates from CMOS processing or nano-imprinting lithography (NIL) can be used without any immediate possibility of degradation. Compared to polymers used in soft-lithography, e.g., PDMS (polydimethylsiloxane) and PMMA (polymethyl methacrylate), Ag$_2$S is relatively harder and therefore issues related to stamp-collapse [11] may not be a serious concern. The lower value of modulus suggests relatively low stresses can be used for the forming process.

**Table 1.** Elastic modulus and hardness of Ag$_2$S compared to commonly used materials in device processing and other materials with comparable mechanical properties.

| Material | Si (100) [12] | Quartz | Au films [10] | Ag films [10] | Sn | **Ag$_2$S** | Pb [13] | PMMA [14] | PDMS [15, 16] |
|---|---|---|---|---|---|---|---|---|---|
| Modulus (GPa) | 169 | 70 | 106 | 85 | 50 | **30** | 16 | 5 | 0.003 |
| Hardness (GPa) | 12.7 | 10 | 2.0 | 1.5 | 0.3 | **0.6** | 0.5 | 0.3 | 0.002 |

*2.2 Elastic Recovery*

To further understand the microforming behavior and quantify elastic recovery of $Ag_2S$, we embossed a commercially available Si grating (from SPI Supplies) with 23±1 nm deep square grooves as shown in figure 2 (top inset). By pressing this grating against the stamp at various stress levels we obtain a curve similar to a stress-stain curve. The actual depth of the embossed pattern was measured using Atomic Force Microscope (AFM) and is plotted as a function of the applied stress in figure 2. Below the yield strength (<80 MPa), some deformation is observed near the periphery of the stamp due to misalignment during pressing. For loads beyond the yield strength, the highest depth observed was about 16±3 nm. Based on these data, an elastic recovery of about 4-10 nm was observed. For higher applied stresses, slight reduction in feature depth is observed which is attributed to strain hardening of the surface.

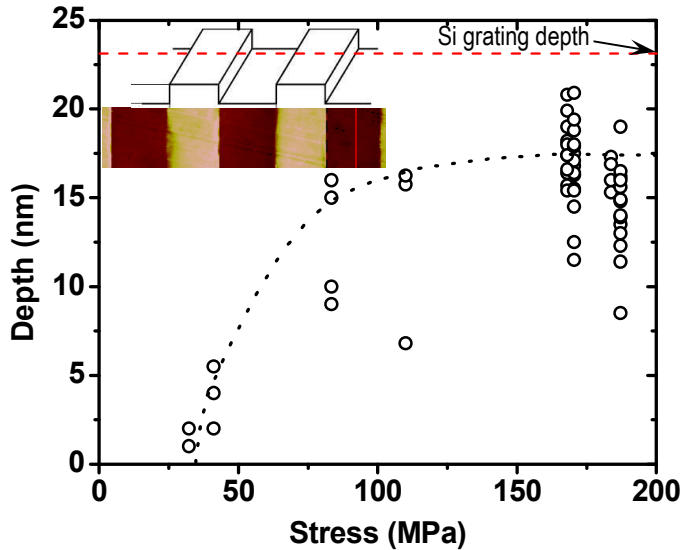

**Figure 2.** Embossed feature depth as a function of applied stress for a Si grating with depth of 23±1 nm (red dash line) and 3 μm pitch as shown in sketch in top inset. Lower inset shows an AFM image of the actual pressed feature on the stamp. Beyond yield strength (80 MPa), elastic recovery of about 4-10 nm is observed which is comparable to nano-indentation results.

*2.3 Material Flow & Pile-up*

Because the embossed stamp is subsequently used for patterning metal films, any deviations of the features on stamp from original template will modify the final transferred pattern. Most important feature modification is related to material flow around the edges (bulging) of the embossed feature. If the mold is shallow such that the bottom of the template presses against the stamp surface, this material flow can be minimized [17].

To estimate the amount of bulging and material pile-up during embossing, we take a two step approach. First, to estimate the upper bound of material flow, we emboss a linear grating with triangular cross-section and very deep features such that no contact of the template bottom to the stamp surface is allowed (top inset, figure 3). A conical indenter—and for the similar regions—a triangular indenter gives higher material pile up around the edges due to one component of the applied stress being normal to the surface [18]. Afterwards, the embossed stamp is pressed against a flat Si surface, a step similar to what is carried out during the etching step in order to estimate the bulging behavior during etching. Because the feature profiles are affected when stamp is brought in contact with metal film, this is a more realistic and closest estimate of how the features appear during etching process.

Figure 3 (lower inset) shows material pile-up around the edges of stamp features as a function of how deep the template was pressed. With AFM, a pile-up of < 20 nm is observed for features as deep as 250 nm. A closer look at the edges shows that most of this pile-up is due to dislocation bursts appearing as shear bands originating at the edges; this height reduces away from the feature edge as observed in the relatively large scattering in the data points. After the stamp is pressed against a flat Si wafer surface with a nominal pressure well below yield strength (~10 MPa), the pile-up reduces to a value <10 nm. After considering the stamp roughness (~3 nm) and the elastic recover (~5 nm), this pile-up is negligibly small.

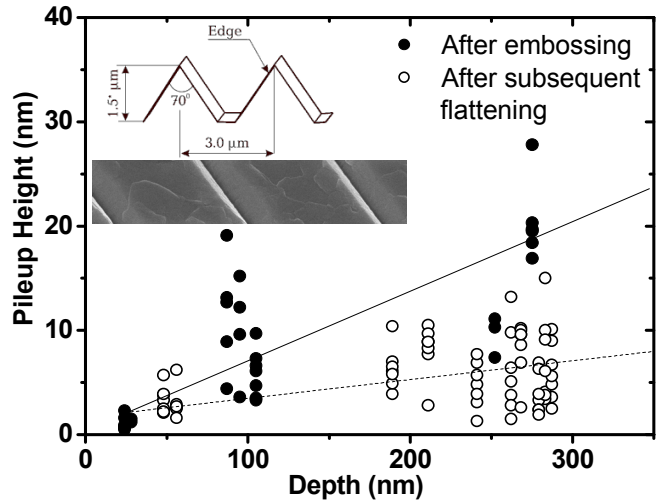

**Figure 3**. Material flow during embossing of a linear grating with triangular cross-section (top inset). SEM image of the stamp surface (lower inset) showing shear bands. The average pile-up (closed circles) and subsequent pressing against a flat Si surface reduces the overall pile-up significantly (open circles).

## 3. Metal patterning

*3.1 Embossing based S4 process*

Figure 4 shows schematic of the embossing based S4 process and SEM images of a bowtie nanoantenna during various steps of the process. For Si template design, bowtie antennas with gaps of 20 nm were first patterned using electron-beam lithography. A 10 nm Cr mask layer was used during standard Reactive-Ion Etching (RIE) (in $CF_4$ atmosphere) to transfer the e-beam pattern onto Si. Although RIE is highly anisotropic, certain amount of side-wall cutting is observed resulting into wider gaps. Additionally, this feature size is close to what e-beam lithography can fabricate with good repeatability. To narrow the gaps, we deposited conformal coating of $Al_2O_3$ using atomic layer deposition (ALD) to fabricate a mean gap size close to 5 nm (figures 4 & 5). Subsequently, this mold was pressed into $Ag_2S$ stamp (at ~3 times the yield strength for 2 minutes) and 20 nm thick Ag films were etched (at 0.4 V and 6 MPa contact pressure). The smallest Ag bowtie gap that could be fabricated using embossed $Ag_2S$ stamps was approximately 8 nm (figure 4 (c)).

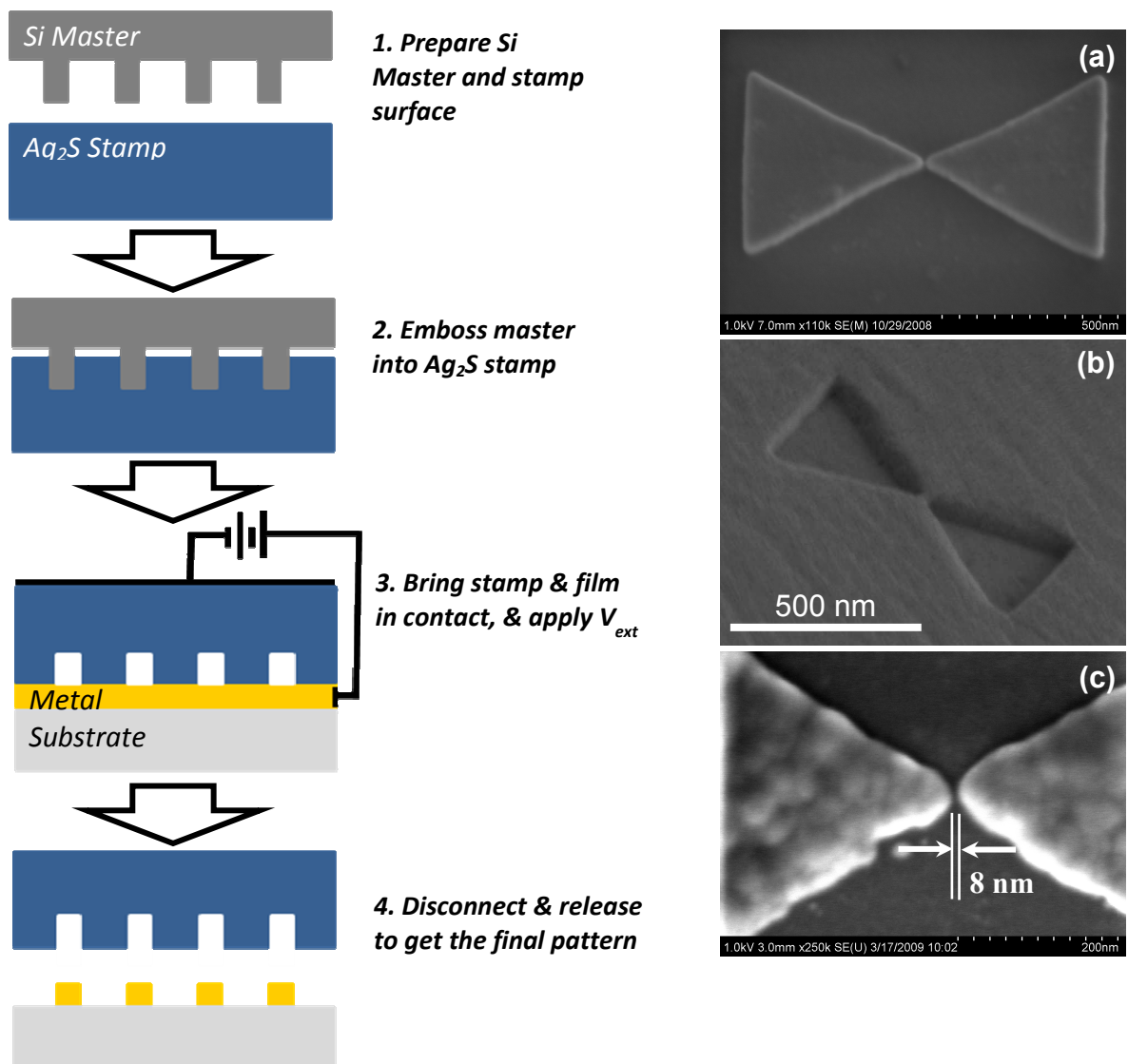

**Figure 4.** (Left) Steps showing the embossing based S4 process. An electron-beam lithography designed Si master is embossed into $Ag_2S$ stamp, which is used for electrochemical etching of metal films. (Right)

Representative images of a bowtie nano-antenna on Si master (a) and resulting feature after embossing into Ag$_2$S stamp (b). Ag bowtie antennas with gaps <10 nm (c), fabricated using an embossed stamps.

To quantify the fidelity loss during pattern transfer, an array of 45 such bowtie gaps were analyzed (figure 5). We observe a mean feature size for the gaps in Si molds close to 5 nm and standard deviation of 2.5 nm. A similar analysis of the Ag bowtie gaps shows this feature size was replicated for about one-third of the bowties suggesting capability to fabricate extremely small features, at rather low yields. Overall, a mean gap size of 18 nm was observed with a standard deviation of 14 nm. Therefore, an approximately 15 nm fidelity loss is observed during two pattern transfers—from Si master to Ag$_2$S stamp to final Ag pattern. An important aspect of this fidelity loss seems to be related to the roughness and grain-size of silver films and possibilities of improvement using smoother Ag films [19] are currently being investigated.

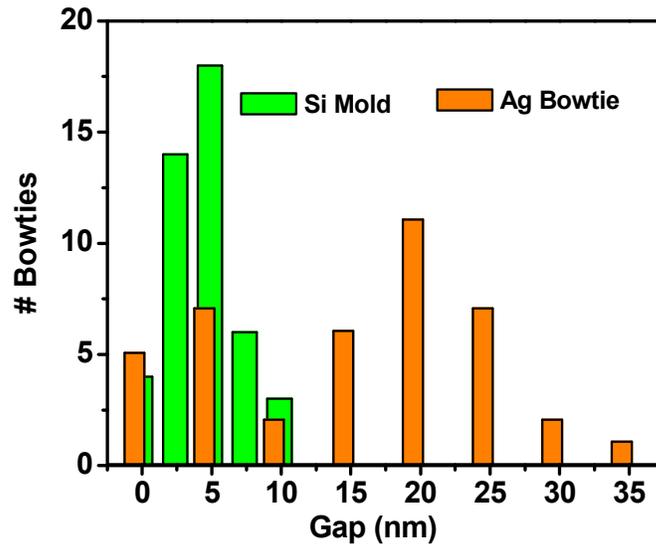

**Figure 5.** Silicon bowtie nanoantennas fabricated using e-beam lithography and post-processed using conformal coating of Al$_2$O$_3$ showed mean gap-size of 5 nm (green). After embossing and electrochemical etching of a 20 nm Ag film, this gap-size was replicated in one-third of Ag bowties; however, overall final mean gap-size was observed to be 18 nm (orange).

Embossing based S4 has additional advantage of being scalable for large area patterning. Due to its low yield strength, Ag$_2$S requires modest pressure for embossing patterns over several square millimeters. As proof of concept, a linear grating (3 μm pitch, 1.5 μm spacing) was pressed into an Ag$_2$S disc of 6 mm diameter and subsequently transferred on to an Ag film (figure 6). For scaling up the process to larger areas issues related to improved conformability, precise alignment and ability to apply higher pressure during pattern transfer need to be addressed and will be reported separately.

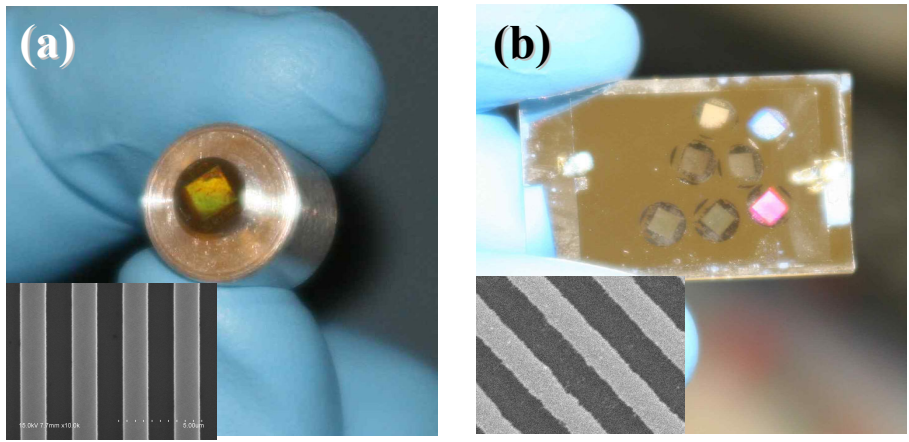

**Figure 6.** Large area patterning using an embossed stamp. (a) shows an Ag$_2$S stamp of 6 mm diameter embossed with a Si linear grating (3 μm pitch and 1.5 μm line-width) while (b) shows patterns transferred onto a 50 nm thick Ag film using this stamp. The insets show SEM images of respective grating patterns.

*3.2 Other advantages of embossing based approach*

An important drawback of indirect patterning processes, e.g., photo-lithography is that control in vertical direction is very limited. However, embossing of solid stamps allows patterning of features with given side-wall angle, and can be faithfully transferred to final metallic features. Additionally, by repeating some of the steps during embossing or etching, complex structures can be fabricated from rather simple templates. Figure 7 (a-b) shows how a linear grating can be used to fabricate two-dimensional structures. In figure 7 (a) a stamp embossed with linear grating was used to first etch a linear pattern while a second etching step with 60° rotation of stamp resulted into an array of rhomboidal microstructures. The secondary step can also be applied during embossing allowing modification of the pattern directly on to the stamp. Figure 7 (b) shows an Ag pattern with different feature heights fabricated by embossing a triangular grating with different pressures and three 120° rotations. The three resulting linear patterns of variable heights can be clearly seen in the inset. In figure 7 (c), a template with conical needles was successfully transferred to demonstrate how embossing allows patterning beyond 2D nanostructures. These approaches were combined to show etching of 800 nm tall Ag square domains with slanted walls by double embossing of a linear grating in figure 7 (d).

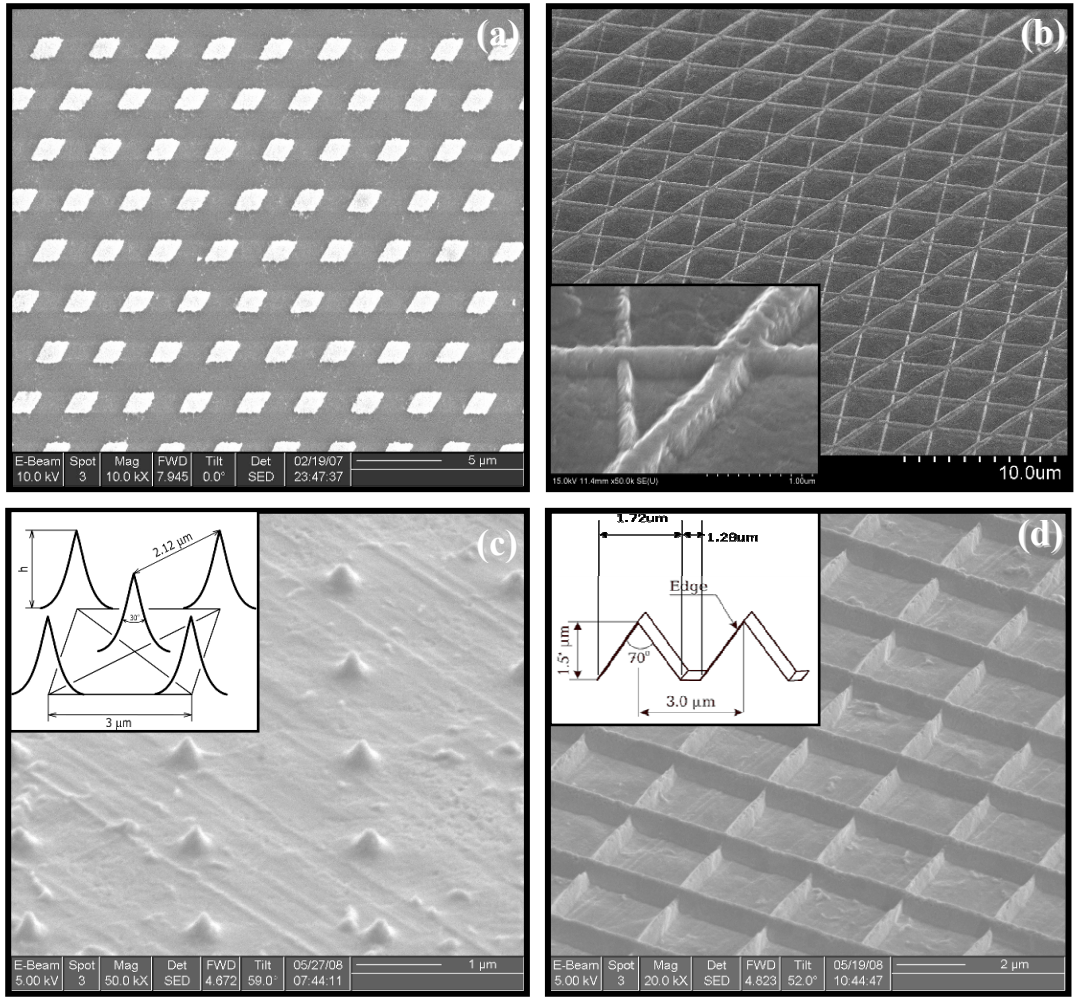

**Figure 7.** Ag patterns that are unique to the embossing based approach: (a) a rhomboidal microstructure array fabricated by double etching of a stamp embossed with linear grating; (b) patterns with variable heights and line widths fabricated in single step by controlling the applied pressure onto a wedge-shaped linear grating shown in inset of (d). The inset shows close up of the lines generated by embossing at three different pressures and rotating the mold every time by 120°; (c) shows a conical pattern (inset) embossed to generate sharp Ag tips with potential applications , e.g., in field-emission displays; (d) shows an 800 nm tall Ag pattern designed by double-embossing a wedge-shaped grating (inset).

## 4. Conclusion

In conclusion, microforming of solid superionic stamps combined with the S4 offers a very competitive process of direct metal patterning with sub-10 nm feature over large areas. The process is highly cost and energy efficient and has ability to integrate with CMOS technology through the use of Si and quartz templates. Our characterization of the stamp material shows issues related to yield strength, material flow, and fidelity loss are not very serious deterrents to scaling the process to commercial scale manufacturing. This direct approach allows side wall control of features in vertical direction, opening new opportunities for patterning slanted features which are beyond the capability of current photolithography processes.

## Acknowledgments

The authors would like to thank Dr. Julio Soares (MRL, UIUC), Andrew Gardner and Prof. William King (both of MechsE, UIUC) for help on nano-indentation experiments. This work was carried out in part in the Frederick Seitz Materials Research Laboratory Central Facilities, University of Illinois, which are partially supported by the U.S. Department of Energy under grants DE-FG02-07ER46453 and DE-FG02-07ER46471.

## References

[1] Madau, M 2002 *Fundamentals of Microfabrication* (New York: CRC Press)
[2] Chou S Y, Krauss P R and Renstrom P J 1997 *J. Vac. Sci. Technol.* B **14** 4129
[3] Rogers J A and Lee H H 2008 *Unconventional Nanopatterning Techniques and Applications* (New Jersey: John Wiley)
[4] Hsu K H, Schultz P L, Ferreira P M and Fang N X 2007 *Nano Lett.* **7** 446
[5] Schultz P L, Hsu K H, Fang N X and Ferreira P M 2007 *J. Vac. Sci. Technol.* B **25** 2419
[6] Wagner C 1953 *J. Chem. Phys.* **21** 1891
[7] Kumar A, Hsu K, Jacobs K, Ferreira P and Fang N 2009 *Mater. Res. Soc. Symp. Proc.* **1156** D07-04
[8] Li X and Bhushan B 2002 *Mat. Charac.* **48** 11
[9] Moser B, Kuebler J, Meinhard H, Muster W and Michler J 2005 *Adv. Mater. Eng.* **7** 388
[10] Caoa Y, Allameh S, Nankivil D, Sethiaraj S, Otiti T and Soboyejo W 2006 *Mater. Sci. Eng.* A **427** 232
[11] Huang Y Y, Zhou W, Hsia K J, Menard E, Park J-U, Rogers J A and Alleyne A G 2005 *Lang.* **21** 8058
[12] Grillo S E, Ducarroir M, Nadal M, Tournié E and Faurie J-P 2003 *J. Phys.* D **36** L5
[13] Almasri A H and Voyiadjis G Z 2010 *Acta Mech.* **209** 1
[14] Briscoe B J, Fiori L and Pelillo E 1998 *J. Phys.* D **31** 2395

[15] Carrillo F, Gupta S, Balooch M, Marshall S J, Marshall G W, Pruitt L and Puttlitz C M 2005 *J. Mat. Res.* **20** 2820

[16] Wu C-L, Lin H-C, Hsu J-S, Yip M-C and Fang W 2009 *Thin Solid Films* **517** 4895

[17] Geiger M, Kleiner M, Eckstein R, Tiesler N and Engel U 2001 *Anna. CIRP* **50** 445

[18] Fischer-Cripps A C 2004 *Nanoindentation* (New York: Spring-Verlag)

[19] Logeeswaran V J, Kobayashi N P, Islam M S, Wu W, Chaturvedi P, Fang N X, Wang S Y and Williams R S 2009 *Nano Lett.* **9** 178